# Frailty Assessment in Aortic Stenosis based on Dynamic Interconnection between Cardiac and Motor Systems

Patricio Arrué[1], Kaveh Laksari[1,2], Nancy Sweitzer[3], Mindy Fain[4,5], Nima Toosizadeh*[1,4,5]

[1]Department of Biomedical Engineering, University of Arizona, Tucson, AZ, United States.
[2]Department of Aerospace and Mechanical Engineering, University of Arizona, Tucson, AZ, United States.
[3]Cardiovascular Division, Department of Internal Medicine, Washington University, St. Louis, MO, United States.
[4]Arizona Center on Aging (ACOA), Department of Medicine, University of Arizona, Tucson, AZ, United States.
[5]Division of Geriatrics, General Internal Medicine and Palliative Medicine, Department of Medicine, University of Arizona, Tucson, AZ, United States.

## Abstract

**Background**: Aortic stenosis (AS) is the most common acquired valvar disease and is associated with increased risk for frailty. Frailty as a geriatric syndrome is associated with muscle weakness and a compromised autonomic nervous system (ANS) performance in older adults. The purpose of the current work was to assess differences in both motor and ANS performance, and interaction between them, as symptoms of frailty in community dwelling older adults with and without AS.

**Methods**: Older adults (≥55 years) with and without AS were recruited. Frailty was assessed using the Fried phenotype. Participants performed an upper-extremity function (UEF) physical task – 20 seconds of rapid elbow flexion of the right arm. Arm motion was measured using gyroscopes and heart rate (HR) was measured using electrocardiogram (ECG) sensor attached on the left side of the chest. Outcomes included UEF motor score (a validated score from 0: "not frail" to 1: "extremely frail" based on slowness, weakness, exhaustion, and flexibility), HR percentage increase due to physical activity and decrease during recovery, and ANS performance (scored from 0: "poor" to 1: "excellent"). ANS performance was measured using convergent cross mapping (CCM), representing interconnection between HR and motor data. ANOVA models were used with the Fried frailty, AS condition, age, BMI, and sex as independent and UEF outcomes as dependent variables.

**Results**: Eighty-six participants were recruited, including 30 with (age=72±11, 10 non-frail and 20 pre-frail/frail) and 56 without AS (age=80±8, 12 non-frail and 44 pre-frail/frail). There was a significant difference in UEF motor score between older adults with and without AS ($p<0.01$, mean values of 0.57±0.25 and 0.48±0.23, respectively). Differences in UEF motor score was also observed between the frailty groups ($p=0.02$, mean values of 0.55±0.24 and 0.40±0.20 for pre-frail/frail and non-frail, respectively). CCM parameters showed significant differences between the frailty groups ($p=0.02$, mean CCM of 0.69±0.05 for non-frail and 0.54±0.03 for pre-frail/frail), but not between the AS groups ($p>0.70$). No significant interaction was observed between frailty and AS condition ($p>0.08$).

**Conclusion**: Current findings suggest that ANS measures may be highly associated with frailty regardless of AS condition. Combining motor and HR dynamics parameters in a multimodal model may provide a promising tool for frailty assessment.

## Introduction

Frailty is a geriatric syndrome associated with loss of physiological reservoir and, in consequence, augmented risk of hospitalization, negative therapy outcomes, disability, and mortality (1). Muscle loss and weakness (sarcopenia and dynapenia) are the main symptoms of frailty (2), caused by inflammatory (elevated interleukin 6 (IL-6), C-reactive protein (CRP), and tumor necrosis factor alpha (TNFα)), metabolic (deficiencies of various mitochondrial subunits), and hormonal derangements (cortisol and testosterone) that shift homeostasis from an anabolic to a catabolic state (3–9). Notably, similar inflammatory, metabolic, and hormonal frailty contributors has been observed in heart diseases, which can further exacerbated by the lack of cardiovascular reserve and a compromised autonomic nervous system (10–15). All these factors can move frail individuals with heart diseases to a more imbalance (less homeostatic) and already stressed state, causing an inability to respond to additional stress, such as therapy burden. Consequently, assessing frailty would be useful for cardiologists as an associated risk score, improving selection of candidates for invasive therapies, by identifying individuals with progressed frailty that may develop therapy complications, as well as those that can reverse their frailty.

Heart diseases are frequent and mortal with nearly 523 million cases worldwide (16). Remarkably, aortic stenosis (AS) is the most common acquired valvar disease, with a prevalence of more than 12% in older adults beyond 75 years old (17). Although much progress has been made in treatment of patients with AS, morbidity, mortality, and the economic burden remain unacceptably high (18–20). For instance, average aggregate 6-month inpatient costs are above $60,000 and mortality is as high as 50% for severe AS, with a predicted doubling AS cases in the next 50 years (21). Like other types of heart disease, AS is a disease of aging and associated with risk for frailty (up to 50% of patients awaiting for AS intervention are frail) (22), and is becoming more frequent as the population average age increases. Frailty assessment is, however, not common in cardiology, especially for AS patients, because current assessment tools are impractical to implement in busy clinical environments (23). While frailty assessments such as Fried frailty phenotype or Rockwood deficit index show promising results in predicting AS therapy outcomes, they are arduous, fully or partially subjective, require trained clinical staff to perform, or require walking for physical assessment (24,25). More importantly, no disease-specific tool is available to identify heart disease-related frailty that can be implemented for AS patients.

We have previously developed a methodology for assessing frailty that incorporates an upper-extremity function (UEF), the corresponding heart rate (HR) response, and the cardiac-motor interconnection (26). The UEF test consists of repetitive and rapid elbow flexion and extension (27), during which several kinematics features representing dynapenia are measured using motion sensors (1). Since UEF involves upper-extremity motion, it is feasible to perform for bedbound patients and when walking tests are difficult for frail older adults. In addition, we showed that HR dynamics, direct measures of sympathetic (HR behavior during activity) and parasympathetic (HR behavior during recovery) performance, were significantly associated with frailty (28). Finally, we validated the cardiac-motor interconnection parameters, generated using convergent cross-mapping between motor function and HR dynamics, as independent predictors for frailty (26). Combining UEF motor, cardiac functions, and their interconnection, we were able to establish a multimodal frailty assessment tool with higher accuracy compared to models including each of the motor or HR dynamics parameters separately (29).

The goal of the current study was to first investigate frailty symptoms in AS patients (motor and cardiac performance) in comparison with non-AS older adults, and second, determine outcome predictive of frailty status regardless of heart disease condition. Our first hypothesis was that both frailty and AS condition would influence motor and HR dynamics parameters in similar ways. Based on our previous research, our second hypothesis was that due to frailty, a weaker interconnection would exist between motor and HR performance that independent of AS condition would be associated with frailty status. Current findings will pave the way to develop a heart disease-specific tool for identifying frailty.

## Methods

*Participants*

Two groups of older adult participants were recruited. Non-aortic stenosis (NAS) older adults (≥55 years) were enrolled between October 2016 and March 2018 from primary, secondary, and tertiary health care settings such as primary and community care providers, assisted living facilities, retirement homes, and aging service organizations. Also, older adults with diagnosed AS (≥55 years) were recruited from the Banner University Cardiovascular program – Advanced Heart Failure and TAVR clinics between September 2021 and October 2022.

The inclusion criteria for the NAS group were: 1) being 55 years or older; 2) the ability to walk a minimum distance of 4.57 m (15 ft) for frailty assessment; and 3) the ability to read and sign an informed consent. For the AS group, an additional inclusion criterion was being diagnosed for aortic stenosis. The exclusion criteria for both groups were: 1) severe motor disorders (Parkinson's disease, multiple sclerosis, or recent stroke); 2) severe upper-extremity disorders (e.g., bilateral elbow fractures or rheumatoid arthritis); 3) cognitive impairment identified by a Mini Mental State Examination (MMSE) score ≤ 23 (30); 4) terminal illness; 5) diseases/treatments that could bias the HR measurements (including arrhythmia and use of pacemaker); and 6) usage of β-blockers or similar medications that can influence HR. Written informed consent was obtained from all participants. The study was approved by the University of Arizona Institutional Review Board. All research was performed in accordance with the relevant guidelines and regulations, according to the principles expressed in the Declaration of Helsinki (31).

*Frailty assessment and clinical measures*

Frailty assessment was executed using the five-component Fried phenotype (1), which includes the following five criteria: 1) unintentional weight loss of 4.54 kg (10 pounds) or more in the previous year; 2) grip strength weakness (adjusted with body mass index (BMI) and sex); 3) slowness based on the required time to walk 4.57 m (15 ft) (adjusted with height and sex); 4) self-reported exhaustion based on a short two-question version of Center for Epidemiological Studies Depression (CES-D); and 5) self-reported low energy expenditure based on a short version of Minnesota Leisure Time Activity Questionnaire (32). Participants were categorized into three frailty groups, which were non-frail if they met none of the criteria, pre-frail if they met one or two criteria, and frail if they met three or more criteria. Collected clinical measures included: 1) MMSE and Montreal Cognitive Assessment (MoCA) for cognition (30,33); 2) comorbidity based on Charlson Comorbidity Score (CCI) (34); 3) depression using Patient Health Questionnaire (PHQ-9) (35); and for AS patients 4) The Kansas City Cardiomyopathy Questionnaire (KCCQ) to measure the quality of life (QoL) (36). Clinical measures with significant association with frailty were considered for both groups as adjusting variables in the statistical analysis because they could potentially influence physical activity and the cardiovascular system performance.

*Upper extremity function (UEF) test*

After completing questionnaires, participants were asked to sit on a chair and rest for two minutes to achieve a normal resting state. Participants then performed the UEF task of elbow flexion-extension as quickly as possible for 20 seconds with the right arm. In a separate study, we showed that UEF results are similar on both left and right hands (37). After the UEF task, participants rested on the chair for another two minutes. Before the test, participants practiced the UEF test with their non-dominant arm to become familiar with the protocol. The protocol was explained to participants and using the exact same verbal instruction they were encouraged only once, before elbow flexion, to do the task as fast as possible. Wearable motion sensors (triaxial gyroscope sensors, BioSensics LLC, Cambridge, MA, sampling frequency=100 Hz; Figure 1A) were used to measure forearm and upper arm motion, and ultimately the elbow angular velocity. Angular velocity data from gyroscopes were filtered using a first-order high pass Butterworth filter with a cutoff of 2.5 Hz. Maximums and minimums of the angular velocity signal were detected, and subsequently, elbow flexion cycles were identified. Motor performance was assessed to represent: 1) slowness based on speed of elbow flexion; 2) flexibility based on range of motion, 3) weakness based on strength of upper-extremity muscles; 4) speed variability and motor accuracy; 5) fatigue based on reduction in speed during the 20-second task, and 6) number of flexion cycles. A sub-score was assigned for each of those features, determined previously based on multivariable ordinal logistic models, with the Fried frailty categories as the dependent variable and UEF parameters plus demographic information as independent variables (28). The normalized UEF motor score from zero (resiliency) to one (extreme frailty) was computed as the sum of sub-scores corresponding to performance results and demographic information (i.e., BMI) (27). More details about UEF validation and the normalized score are explained in previous research (27,37,38).

*Heart rate (HR) measures*

HR was recorded using a wearable ECG device with two electrodes and one built-in accelerometer (360° eMotion Faros, Mega Electronics, Kuopio, Finland; ECG sampling frequency=1000 Hz and accelerometer sampling frequency=100 Hz; Figure 1A). One ECG electrode was placed on the upper mid-thorax and the other one inferior to the left rib cage. The placement of the electrodes on the left chest would minimize the movement artifacts due to UEF test with the right arm. ECG data was analyzed for 20 seconds of baseline, 20-second UEF, and 30 seconds of recovery. We used accelerometer data within the ECG sensor to manually select the start and end points of the UEF test. Since the accelerometer and ECG data are synchronized, the start and end points were used as time stamps for indicating baseline, task, and recovery periods. Once the task start point is selected, 20 seconds of baseline HR data was added to create the baseline period. Similarly, for the 30 second recovery period the end point of the activity was used to add 30 seconds of HR data. RR intervals (successive R peaks of the QRS signal) were computed using the Pan-Tompkins algorithm (39). The automated peak detection process was manually inspected by two researchers (PA and NT). Two types of HR parameters were extracted, one representing baseline HR and HR variability (HRV), and one representing HR dynamics (changes in HR during UEF and HR recovery after the task) (28,40,41). HR dynamics included time to reach maximum peak and recovery HR, as well as percent increase and decrease in HR during activity and recovery periods, respectively. HR time to peak was defined as the required time to reach maximum HR during the task with reference to minimum HR during the baseline. Similarly, recovery time was defined as the required time to reach minimum HR during the recovery with reference to maximum HR during the task. HR increase was defined as the amount of increase in HR during the task compared to minimum baseline HR, which was presented as the absolute value (bpm) and the percentage of change with respect to minimum baseline HR. Similarly, decrease in HR during the recovery compared to maximum HR during the task was considered as the amount of HR decrease and was presented as the absolute value (bpm) and the percentage change with respect to maximum HR during the task.

*Convergent cross mapping (CCM) analysis*

We quantified the directional nonlinear interconnections between HR and motor data using CCM. An overview of CCM is presented in Figure 1. This method uses a historical trace of HR for predicting motor performance (or inversely, a historical trace of motor performance for predicting HR). Firstly, we created evenly sampled data of synchronized HR and motor function with a sampling frequency of 10Hz, using spline interpolation (Figure 1B). Each HR data point represents average HR values over 0.1 seconds. Corresponding motor data represent the angular displacement travelled during each 0.1 second of UEF. For calculating motor performance, motor function $M_f$ was defined by:

$$M_{f_i} = \int_{t_i}^{t_i+0.1} \omega_e dt, \tag{1}$$

where $\omega_e$ represents the rectified angular velocity of the elbow.

Takens's embedding theorem was used here, which guarantees that the information of a chaotic dynamic system could be represented from a single-observed time series $X$ as an $E$-dimensional manifold (42). The shadow manifold $M_X$ consists of an $E$-dimensional data with lagged coordinates ($\tau$) of the variable:

$$M_X = \langle X(t), X(t-\tau), X(t-2\tau) \ldots X(t-(E-1)\tau) \rangle. \tag{2}$$

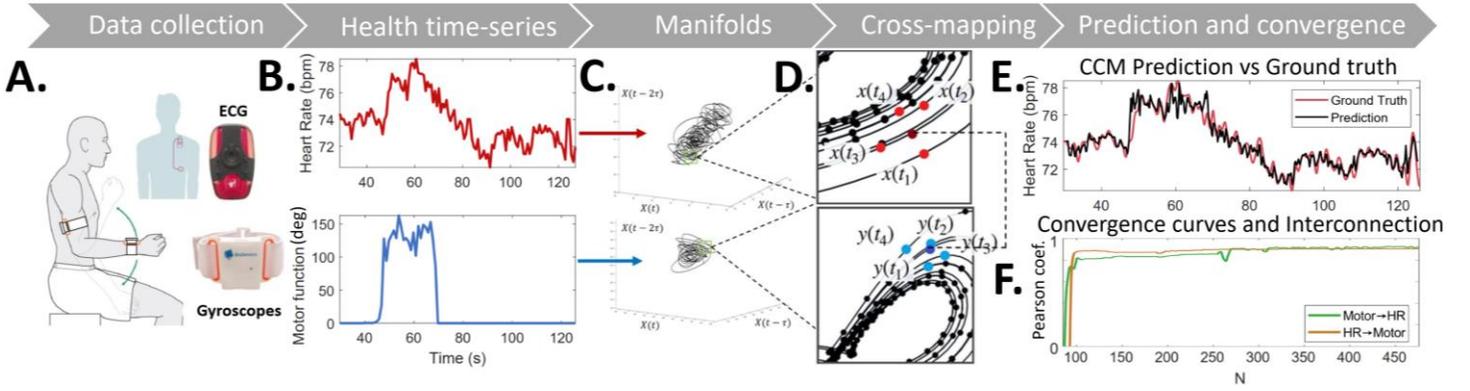

**Figure 1:** Overview of the CCM method to assess interconnection between motor and HR data: **A.** wearable devices (gyroscopes) to obtain angular velocity and ECG during the UEF physical task; **B.** motor performance and HR extraction; **C.** CCM shadow attractor manifolds on time-lagged coordinate systems; **D.** prediction of HR from motor function and vice-versa in a time point (dark red and dark blue dots, respectively) using a distance-based weighted average of neighbors (bright red and bright blue dots); **E.** comparison between predicted HR data and ground truth, also applicable to motor function; and **F.** convergence curves of Pearson correlation coefficient between predicted and ground truth as a function of library length (data points used for developing manifolds).

We built $E$-dimensional manifolds from each of these two time series (42) (Figure 1C). Dimension ($E$) of four was selected based on the average false nearest neighbor approach (43). A time lag ($\tau$) of one second was used for analysis based on the delayed mutual information method (44). We predicted one time series (e.g., HR) by historical records of the other signal (e.g., shadow manifold of motor function) using a k-nearest neighbor technique. For a dimension $E$, we determined $E+1$ nearest neighbors and identified indices of each data points in manifolds ($M_X$). Using these indices for one manifold (e.g., HR $X(t)$), we found corresponding neighbors in the second manifold (e.g., motor function data $Y(t)$) (Figure 1D), and then predicted $\hat{Y}(t)$ from $X(t)$ as the weighted mean of $E+1$ points in the second manifold (45):

$$\hat{Y}(t) = \sum_{i=1}^{E+1} w_i Y(t_i), \qquad (3)$$

where $w_i$ weights are calculated based on the Euclidean distances between $M_Y$ and its $i^{th}$ nearest neighbor on $X(t)$. The Pearson correlation coefficient between the predicted and original time series was calculated to assess the strength of interconnections between motor and HR data (Figure 1E). As documented in previous studies, the correlation coefficient is expected to increase with increasing the time-series length (i.e., library length, Figure 1F). We selected CCM parameters corresponding to correlation values at the maximum library length. The CCM calculations were performed using MATLAB (R2022a, MathWorks, Natick, MA, USA).

*Statistical analysis*

We used Analysis of variance (ANOVA) models to determine the differences in demographic information between frailty groups, except for sex. Instead, the chi-square ($\chi^2$) test was used to assess the difference in sex categories between groups, as well as differences in frailty categories between AS vs. NAS groups. UEF, HR dynamics, and CCM parameters were compared between frailty groups and NAS/AS conditions using multivariable ANOVA models; age, sex, and BMI were considered as adjusting variables since they have been previously associated with motor and cardiac performance and frailty (27,46–48). Cohen's effect size (*d*) was estimated. ANOVA analyses for comparing motor, cardiac, and CCM parameters across frailty groups and NAS/AS conditions were repeated with clinical measures with significant association with frailty as covariates. Effect sizes are interpreted as small (<0.2), medium (>0.2 and <0.5), and large (>0.8) (49). The statistical analysis was performed using JMP® Pro (Version 16, SAS Institute Inc., Cary, NC, USA).

## Results

*Participants and clinical measures*

Fifty-six NAS older adult participants were recruited for the study, including 12 non-frail (age=76.92±7.32 years), 40 pre-frail (age=80.53±8.12 years), and four frail individuals (age=88.25±4.43 years). On the other hand, 30 AS patients were enrolled, including 10 non-frail (age=69.00±3.61 years), 17 pre-frail (age=72.71±2.77 years), and 3 frail individuals (age=75.67±6.60 years). Of note, due to the small number of frail participants (n=7), frail and pre-frail groups were merged for the statistical analysis. A summary of demographics is presented in Table 1. There were significant differences in age and height between AS and NAS groups, and therefore all statistical analyses were adjusted for these two variables.

*Effect of frailty*

There was no significant difference in frailty categories between AS and NAS groups. Significant effects of frailty on UEF motor score, resting HR baseline, HR percentage increase and decrease, and CCM values were observed across the participants, as reported in Table 2 and Figures 2.A, 2.C, 2.E, 2.G, 2.I and 3.A. Pre-frail/frail older adults showed higher UEF motor score, higher baseline HR (in rest), lower changes in HR due to physical task, and smaller CCM, compared to non-frail older adults.

**Table 1.** Demographic information and clinical measures of participants.

| Variables | Non-frail (n=22) | Pre-frail/Frail (n=64) | *p*-value (effect size) | Non-aortic stenosis (n=55) | Aortic stenosis (n=31) | P-value (effect size) |
|---|---|---|---|---|---|---|
| Female, n (% of the group) | 12 (54.55) | 42 (65.63) | 0.49 | 40 (72.73) | 14 (27.27) | 0.02* |
| Age, year (SD) | 73.31 (9.17) | 78.70 (10.12) | 0.06 (0.48) | 80.09 (8.02) | 72.42 (11.63) | <0.001* (0.77) |
| Height, cm (SD) | 166.31 (11.17) | 166.21 (10.25) | 0.81 (0.06) | 164.24 (10.05) | 169.77 (10.29) | 0.02* (0.55) |
| Weight, kg (SD) | 72.30 (16.85) | 77.02 (20.95) | 0.26 (0.28) | 73.59 (19.04) | 79.75 (21.34) | 0.14 (0.34) |
| Body mass index, kg/m$^2$ (SD) | 26.12 (5.63) | 27.59 (5.85) | 0.29 (0.27) | 27.08 (5.82) | 27.46 (5.83) | 0.68 (0.09) |
| CCI score, 0-29 (SD) | 4.45 (1.84) | 5.80 (2.39) | <0.01* (0.68) | 5.09 (2.12) | 6.10 (2.56) | 0.02* (0.53) |
| **Fried criteria, n (% of the group)** | | | | | | |
| Non-frail subjects | - | - | - | 12 (22%) | 10 (32%) | 0.29 ( - ) |
| Weight loss | 0 | 3 (5%) | - | 1 (2%) | 2 (6%) | - |
| Weakness | 0 | 27 (42%) | - | 17 (31%) | 10 (32%) | - |
| Slowness | 0 | 41 (%) | - | 34 (62%) | 7 (23%) | - |
| Exhaustion | 0 | 18 (%) | - | 7 (13%) | 11 (35%) | - |
| Low energy | 0 | 13 (%) | - | 8 (15%) | 5 (16%) | - |

*Effect of AS condition*

As reported in Table 2, significant effect of AS condition (AS vs. NAS) on UEF motor score and HR percentage increase and decrease were observed (*p*<0.01, Figures 2.B, 2.D and 2.F), as well as time to peak and HR recovery time (*p*<0.0001, Figures 2.L and 2.N). AS individuals showed worse UEF motor score and larger (in percentage) but slower HR changes due to the physical task. In contrast to UEF motor score and HR dynamics results, which showed significant differences between both frailty groups and the AS condition, CCM parameters were only significantly different across frailty groups (*p*=0.04 and <0.01, Figure 2.G and 2.I), and not the AS condition (*p*>0.62, Figure 2.H and 2.J, and Table 2).

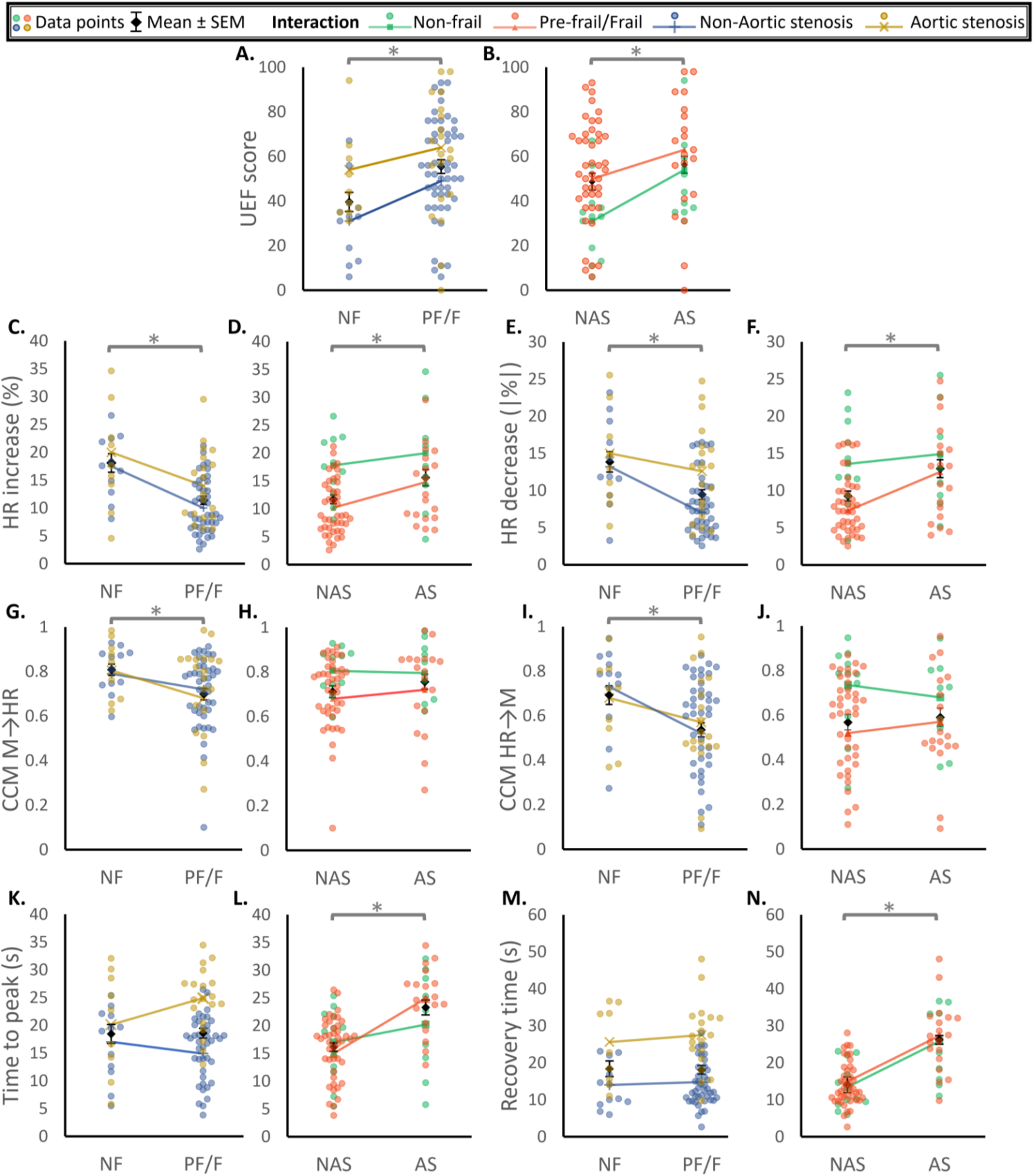

**Figure 2.** UEF score (**A.** and **B.**), HR percentage changes (**C.** to **F.**), CCM parameters (**G.** to **J.**), time to HR peak from rest (**K.** and **L.**) and time to recover to HR baseline (**M.** and **N.**) for frailty and aortic stenosis categories. Each data point corresponds to an individual. Interaction profiles are incorporated to show the cross-effects between different subgroups. A significant between group difference is identified by the asterisk (*p*<0.05).

**Table 2.** Differences in UEF, HR dynamics, and CCM parameters across frailty groups. A significant association is represented by the asterisk.

| Parameters | Non-frail (n=22) | Pre-frail/Frail (n=63) | P-value (Effect size) | Non-aortic stenosis (n=55) | Aortic stenosis (n=30) | P-value (Effect size) |
|---|---|---|---|---|---|---|
| **UEF motor score** | | | | | | |
| UEF motor score, 0-100 (SD) | 39.59 (20.20) | 55.46 (24.12) | 0.01* (0.66) | 48.69 (23.43) | 56.23 (24.92) | <0.01* (0.70) |
| **HR dynamics parameters** | | | | | | |
| HR rest mean, bpm (SD) | 68.21 (10.46) | 76.31 (15.99) | 0.01* (0.66) | 76.20 (15.34) | 70.45 (14.27) | 0.24 (0.30) |
| Time to peak HR, seconds (SD) | 18.43 (7.86) | 18.57 (6.85) | 0.71 (0.10) | 16.12 (5.69) | 23.28 (7.19) | <0.0001* (1.28) |
| Recovery time, seconds (SD) | 18.36 (9.71) | 18.07 (9.19) | 0.53 (0.17) | 14.04 (5.83) | 26.20 (9.54) | <0.0001* (1.73) |
| HR percent increase, % (SD) | 18.10 (7.47) | 11.40 (5.63) | <0.0001* (1.10) | 11.71 (5.78) | 15.59 (7.75) | 0.01* (0.62) |
| HR percent decrease, % (SD) | -13.85 (6.12) | -9.45 (5.17) | <0.01* (0.77) | -9.26 (4.95) | -12.96 (6.35) | 0.01* (0.67) |
| HR absolute increase, bpm (SD) | 12.65 (4.05) | 10.42 (8.06) | 0.44 (0.21) | 9.66 (3.89) | 9.61 (4.45) | 0.77 (0.08) |
| HR absolute decrease, bpm (SD) | -10.64 (4.88) | -7.79 (4.34) | 0.06 (0.53) | -7.67 (4.03) | -10.04 (5.31) | 0.06 (0.49) |
| **CCM parameters** | | | | | | |
| Correlation Motor→HR (SD) | 0.81 (0.11) | 0.70 (0.20) | 0.04* (0.55) | 0.71 (0.19) | 0.76 (0.18) | 0.62 (0.13) |
| Correlation HR→Motor (SD) | 0.69 (0.19) | 0.54 (0.25) | <0.01* (0.72) | 0.57 (0.26) | 0.59 (0.22) | 0.72 (0.09) |

*UEF: upper extremity function; SD: standard deviation; HR: heart rate; bpm: beats per minute; CCM: convergent cross-mapping.*

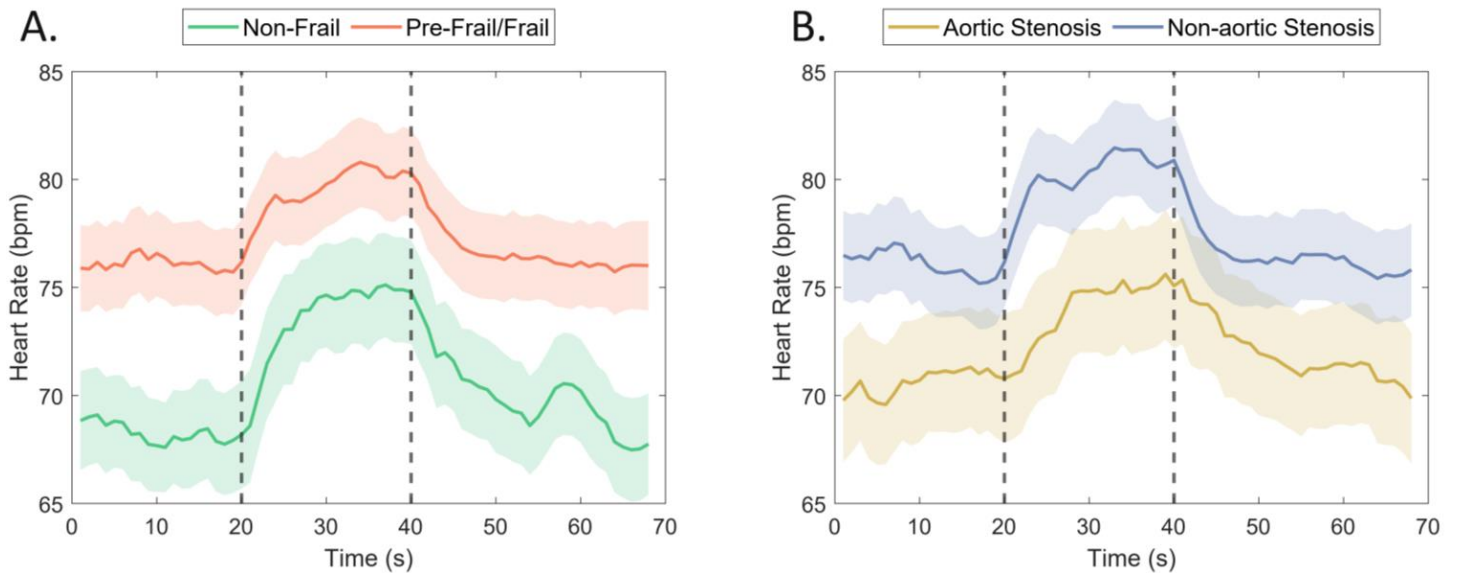

**Figure 3.** Heart rate distributions for frailty (A) and aortic stenosis (B) groups. Means and standard error are represented by solid lines and shaded regions, respectively. Physical task 20-seconds arm test starting and ending points are indicated with vertical lines. Linear interpolation was used to provide equidistant HR time series across samples.

## Discussion

*Effect of frailty on HR dynamics and motor performance*

In confirmation of previous work (40), there were significant associations between frailty status and the level of HR changes during physical activity and the consecutive recovery period. During activity there is an increase in sympathetic outflow, and consequently an increase in HR and stroke volume to satisfy circulatory requirements (50). Afterwards in the recovery period, the parasympathetic activity is enhanced to reduce HR to the baseline level (51–53). An impaired ANS performance is represented by an absence of resilience in changing HR, and previous studies have shown the effect of frailty on ANS outflow. In a study by Weiss et. Al, HR was monitored during a seated step test (54), and in another study by Romero-Ortuno et. al, HR was measured during and after 3 minutes of a lying to standing orthostatic test (55), and they confirmed an impaired HR response due to frailty that was observed by HR changes during the test. In confirmation of these findings, we observed significant lower HR changes in pre-frail/frail population in comparison to non-frail individuals (Figure 2.C and 2.E, Table 2). Our findings suggest, unlike the amount of change, the timing of HR changes (i.e., HR rest-to-peak and recovery times) was not associated with frailty ($p>0.53$).

Motor performance is impacted by frailty through sarcopenia and dynapenia (diminished muscle mass and performance), leading to poorer physical activity (1). Dynapenia is typically quantified by walking speed or grip strength tests. Nevertheless, performing walking tests in clinical settings is challenging, and many frail older adults, especially those with AS may have walking disabilities. Grip strength, on the other hand, only measures muscle strength and cannot reveal other aspects of motor deficits, such as flexibility. We have previously validated the sensor based UEF motor task to detect with an accuracy of 0.7 (27) systematic decrements in motor function associated with frailty, including slowness, weakness, inflexibility, fatigue, and motor variability (38,56). Here, we found significant increases in the UEF motor score for pre-frail/frail participants compared to non-frails (Figure 2.A, Table 2), confirming previous findings. In addition, we observed that frail participants with additional burden of AS showed even higher UEF motor score than AS non-frail or NAS frail participants, revealing the accentuated deficits in motor function due to doubling effects of heart disease and frailty. These findings demonstrate the importance of frailty assessment in AS patients who have less physiological reservoirs and an impaired ANS performance, and consequently higher risk of poorer outcomes after invasive therapies such as open-chest surgeries.

*Effect of AS on HR dynamics and motor performance*

According to literature, AS impacts cardiac and motor performances. Historically, HR variability (HRV) has been used for assessing ANS dysfunction and proposed as a vital sign (57). However, a low correlation has been reported between several time and frequency-domain HRV parameters and AS severity (58). In the current work we have found several HR dynamics parameters that could potentially be used for assessing cardiac autonomic dysfunction due to aortic stenosis, such as significantly higher HR changes during and post physical activity, together with a significantly slower HR response to the physical task. A previous study, in which HR was monitored before and after one minute of supine bicycle exercise, showed that HR increase in AS patients were slightly greater than NAS (59). Here, we observed that monitoring HR during the physical activity period provides similar trends. This finding is consistent with other reports (60,61), within which myocardial work, as a measure of oxygen demand defined by the product of myocardial wall stress and wall stress rate, was greater in AS than controls. This observation suggests existence of a relative supply/demand imbalance in AS, which would make these patients more vulnerable to myocardial ischemia (59). In our study we found that AS participants had significantly higher changes in HR, during both physical activity and rest than NAS participants (Table 2, Figure 2.D and 2.F). We attributed the higher HR change in AS to the ANS response to a higher myocardial oxygen need, a consequence of myocardial hypertrophy AS-associated (62). This fact is supported by other work claiming that the rapid rise of HR in AS patients is a compensatory mechanism to maintain cardiac output (63). The aforementioned myocardial ischemia risk and the impaired perfusion to myocytes usually lead to AS patients gradually decrease their physical activity to accommodate their condition (64). This fact is also supported by our results, since the motor performance in AS individuals was significantly worse compared to NAS participants (Table 2, Figure 2.B). Noteworthy, we observed significant differences in HR percentage changes, but not absolute changes, although the trend of higher HR decrease among AS compared to NAS was still observable in absolute HR values. This may occur because of differences in baseline HR data between two groups as demonstrated in Figure 3.B. Nevertheless, more confirmatory investigations in larger samples and age-matched groups of AS and NAS are required in future.

*CCM parameters for identifying frailty*

We previously proved that CCM was an independent predictor for frailty on a community dwelling population, finding a significantly weaker interconnection between motor and cardiac systems for pre-frail and frail participants compared to non-frails (26). We attributed these results to aging-related physiological dysregulation in the motor cortex-reflexes feedback system (65–71), and loss of homeostasis and heightened inflammatory state (3,5–7,9,72). In the current study, CCM parameters showed significant differences across frailty groups for the combined sample of AS and NAS (Figure 2.G and 2.I, Table 2), while similar magnitudes were calculated for NAS vs AS samples for CCM M→HR (p=0.62) and CCM HR→M (p=0.72). These results indicate that CCM parameters are mostly not affected by the AS condition, as indicated in Figure 2.G and Figure 2.I. This observation support the importance of extracting parameters associated with physiological interconnections for a proper and accurate frailty assessment, regardless of comorbid condition (73–75). Findings suggest that CCM provides promising measures to develop an accurate tool for AS-related frailty assessment.

*Limitations and Further work*

Despite the promising findings of the current study, there are some limitations related to the recruited sample. First, there was a limited number of frail participants, and therefore, pre-frail and frail groups were merged. Second, there was no significant difference in the number of frail population between AS and NAS groups (Table 1, p=0.29), which may suggest that the AS sample in the current study may not perfectly represent the prevalence of frailty in this population. Third, age, sex and height were significantly different between our AS and NAS populations. Although we adjusted statistical analyses with age, sex, and BMI, our findings would benefit in future investigation by collecting data from age, sex, and BMI-matched groups. Fourth, participants with arrhythmia and those who require β-blockers and pacemakers were excluded from the study. Also, test-retest reliability of CCM parameters were not investigated here. Therefore, the interconnection analysis should be confirmed in larger studies incorporating test-retest reliability measures. Additionally, we used time-series library lengths that may not provide accurate results for some participants, since some HR data may have a higher level of short-term complexity, leading to less dense attractor shadow manifolds and consequently a non-completely developed convergence of CCM parameters. Possible solutions would be to perform longer arm tests; however, this will lead to more physical demand on frail and those with aortic stenosis.

*Conclusions and clinical implications*

In the present work a novel quantification of interconnection between motor and cardiac autonomic systems was implemented for frailty assessment. We demonstrated that CCM parameters showed weaker interconnection between motor and cardiac systems among pre-frail/frail older adults compared to non-frails, regardless of AS condition. The simplicity of the investigated UEF test permits performing it even for hospitalized bed-bound patients with heart disease, for predicting therapy complications, in-hospital outcomes, and rehabilitation strategies. We expect to develop a heart disease-specific tool for frailty assessment using the CCM parameters, contributing to early assessment and better therapy strategies for AS patients. Further, commercialized wearable devices are now allow accurate assessment of HR and motion. The UEF test required only 20 seconds of assessment and results can be provided in one minute. Showing the proof of concept in the current study, in our future investigation, we will develop an easy-to-use app for Smart Watch for identifying frailty using simultaneous measures of motor and cardiac functions.

# Declarations section

## Ethics approval

The studies involving humans were approved by University of Arizona Institutional Review Board (approved IRB ID: 2105776487). The studies were conducted in accordance with the local legislation and institutional requirements. The participants provided their written informed consent to participate in this study.

## Consent for publication

Not applicable.

## Data availability

The datasets used for the current study are available from the corresponding author on reasonable request.

## Competing interests

The authors declare that they have no competing interests.

## Funding

This project was supported by three awards from the National Institute of Aging (NIA/NIH - Phase 2B Arizona Frailty and Falls Cohort 2R42AG032748–04, NIA/NIH - 1R21AG059202-01A1, and NIH/NIA - 1R01AG076774-01A1) and an award from NSF (NSF 2236689 – CAREER). The views represented in this work are solely the responsibility of the authors and do not represent the views of NIH or NSF.

## Authors' contributions

PA: study design, data collection, data analysis and writing of the paper. KL: study design and contribution to writing. NS: study design and contribution to writing. MF: study design and contribution to writing. NT: study design, data analysis, and writing of the paper. All authors contributed to the article and approved the submitted version.

## Acknowledgements

We want to thank Allison Klatt and Sarver Heart Cerner for data collection.